\documentstyle[prd,aps,floats]{revtex}

\begin{document}
\preprint{}
\draft

%
%
\input epsf
\renewcommand{\topfraction}{0.99}
\twocolumn[\hsize\textwidth\columnwidth\hsize\csname 
@twocolumnfalse\endcsname

\title{Supersymmetry and primordial black hole abundance constraints.} 
\author{Anne M.~Green}
\address{Astronomy Unit, School of Mathematical Sciences, Queen Mary and
Westfield College,\\ Mile End Road, London, E1 4NS,~~U.~~K.}
\date{\today} 
\maketitle
\begin{abstract}
We study the consequences of supersymmetry for primordial black hole
(PBH) abundance constraints. PBHs will emit supersymmetric particles
throughout their evaporation if their mass is less than about
$10^{11}$g. In most models of supersymmetry the lightest of these
particles, the lightest supersymmetric particle (LSP), is stable and
will hence survive to the present day. We calculate the limit on the
initial abundance of PBHs from the requirement that the present day
LSP density is less than the critical density. We apply this limit,
along with those previously obtained from the effects of PBH
evaporation on nucleosynthesis and the present day density of PBHs, to
PBHs formed from the collapse of inflationary density perturbations in
the context of supersymmetric inflation models. If the reheat
temperature after inflation is low, so as to avoid the overproduction
of gravitinos and moduli then the lightest PBHs which are produced in
significant numbers will be evaporating around the present day and
there are therefore no constraints from the effects of the evaporation
products on nucleosynthesis or from the production of LSPs. We then
examine models with a high reheat temperature and a subsequent period
of thermal inflation.  In these models avoiding the overproduction of
LSPs limits the abundance of low mass PBHs which were previously
unconstrained. Throughout we incorporate the production, at fixed
time, of PBHs with a range of masses, which occurs when critical
collapse is taken into account.
\end{abstract}

\pacs{PACS numbers: 98.80.-k, 14.80.Ly \hspace*{6.0cm}  astro-ph/9903484}

\vskip2pc]

\section{Introduction}
Primordial black holes (PBHs) may form in the early universe via a
number of mechanisms~\cite{dpcoll,form}, the simplest of which is the
collapse of large density perturbations~\cite{dpcoll}. Due to quantum
effects PBHs evaporate, mimicking the emission from a black body with
finite size and temperature $T_{{\rm BH}}$ where
\begin{equation}
\label{tbh}
T_{{\rm BH}}= \frac{ \hbar c^3}{8 \pi G M_{{\rm BH}}} = 1.06 \left( 
          \frac{10^{13}}
      {M_{{\rm BH}}} \right) {\rm GeV} \,,
\end{equation} 
and $M_{{\rm BH}}$ is the PBH mass in grams. The standard picture of PBH
evaporation is that all particles which appear elementary at the
energy scale of the PBH and have rest mass less than $T_{{\rm BH}}$
are emitted directly~\cite{MacG}. For instance PBHs with $T_{{\rm
BH}}$ above the QCD quark-hadron transition scale $\Lambda_{{\rm QH}}
\approx 250-300$ MeV, emit relativistic quark and gluon jets which then
fragment into photons, leptons and hadrons as in high energy
accelerator collisions. 

The mass loss,
in grams per second, is given by
\begin{equation}
\label{dm}
\frac{{\rm d} M_{{\rm BH}}}{{\rm d}t} = -5.34 \times 10^{25} f(M_{{\rm BH}})
             M_{{\rm BH}}^{-2} \,,
\end{equation} 
where $f(M_{{\rm BH}})$ is a function of the number of species emitted
and is normalised to 1 for PBHs with $M_{{\rm BH}} \gg 10^{17}$g which
emit only massless particles (see ref.~\cite{MacG} and
Sec.~\ref{sec:cond} for more details).  The effects of the evaporation
products allow constraints to be put on the initial abundance of PBHs
over a range of masses. PBHs with mass in the range $10^{9}{\rm g} <
M_{{\rm BH}}< 10^{13}{\rm g}$, would have evaporated after
nucleosynthesis and could have a number of effects on the successful
predictions of nucleosynthesis~\cite{bnuc} whilst PBHs with $M_{{\rm
BH}} \sim 5 \times 10^{14}$g are evaporating today and the number
density of photons produced must not exceed the observed $\gamma$-ray
background~\cite{bgam}. In the case of lighter PBHs with
$T_{{\rm BH}} > 100$ GeV the fundamental particles emitted, and hence
$f(M_{{\rm BH}})$, depend upon the particle physics model assumed.

 Currently the widely accepted extension of the Standard Model is
supersymmetry~\cite{susy}, where each Standard Model particle has a
supersymmetric partner known as a sparticle. Motivated mainly as an
attempt to understand why the weak scale is much smaller than the
Planck scale (known as the gauge hierarchy problem~\cite{gh}),
supersymmetry also leads to the unification of gauge couplings at an
energy of about $10^{16}$ GeV~\cite{unif}. The phenomenology of the
sparticles is complicated being governed by up to 105 independent and
unknown parameters. In the simplest models the sparticles have masses
of order 100 GeV and there is a multiplicatively-conserved quantum
number known as R-parity, where Standard Model particles have R$=+1$
and sparticles have R$=-1$. Consequentially heavier sparticles decay
into lighter sparticles and the Lightest Supersymmetric Particle (LSP)
is stable, since it has no allowed decay mode. In most models the LSP
is non-relativistic at freeze out and is therefore a candidate for the
cold dark matter (CDM)~\footnote{In gauge-mediated models of
supersymmetry breaking the LSP could be a gravitino with mass $\sim 1$
keV which would constitute warm dark matter~\cite{wdm}.}.

The LSP must be neutral and weakly interacting~\cite{neutweak} as
otherwise it would have condensed, along with the baryonic matter,
into astrophysical structures and the resultant abundance of anomalous
heavy isotopes would exceed observational limits~\cite{abund}. The LSP
may therefore be a sneutrino or the gravitino but in most
supersymmetric theories it is the lightest neutralino $\chi$, which
is a mix of the supersymmetric partners of the photon, the Z boson and
the neutral Higgs boson. Throughout this paper we therefore use
$\chi$ to denote the LSP.

Purely experimental searches at LEP have led to the limit $m_{\chi}
\geq 30$ GeV~\cite{m1}. This limit can be tightened to $m_{\chi}>
42$ GeV by making various theoretical assumptions and requiring that
the present day LSP density lies in the interesting range for CDM:
$0.1 < \Omega_{\chi} h^2 < 0.3$~\cite{m2}. The LSP comoving number
density has remained constant since annihilations ceased at the freeze
out temperature $T_{{\rm f}}$. This leads to a simple estimate (see
Ref.~\cite{susydm} for a review of this and more detailed
calculations) of the current LSP density:
\begin{equation}
\Omega_{\chi} h^{2} = \frac{10^{-3}}{\langle \sigma_{{\rm ann}}(\chi \chi)
         v_{{\chi}} \rangle T_{0} m_{{\rm Pl}}} \,,
\end{equation}
where $\langle \sigma_{{\rm ann}}(\chi \chi) v_{{\chi}} \rangle$ is
the thermally averaged annihilation cross section. As $m_{\chi}$
increases $\langle \sigma_{{\rm ann}}(\chi \chi) v_{{\chi}} \rangle$
decreases so that $\Omega_{\chi} h^2$ increases. This leads to an upper
limit of $m_{\chi} \leq 300$ GeV~\cite{m3}, or $m_{\chi} \leq 600 $GeV if
co-annihilations with the stau slepton, which are important in some
regions of parameter space, are taken into account~\cite{m4}.

Supersymmetric particles which are produced by the evaporation of PBHs
after the temperature of the universe has fallen below $T_{{\rm f}}$
will not be able to equilibrate due to the inefficiency of
annihilations. In Sec.~\ref{sec:cond} we calculate the mass range of
PBHs which are heavy enough to evaporate after LSP freeze out but
light enough to produce sparticles. The LSPs produced by the decay of
these sparticles, along with the LSPs evaporated directly, will
therefore provide an additional contribution to the present day
density of LSPs. Whilst it is possible that the number density of LSPs
produced via PBH evaporation may be comparable to the freeze out
number density, reducing the upper limit on $\Omega_{\chi}$ and hence
$m_{\chi}$, this would require extreme fine tuning of the initial
abundance of PBHs. We therefore use the conservative requirement that
the present day density of LSPs produced via PBH evaporation after
$T_{f}$ is less than the critical density, to constrain the initial
abundance of LSP producing PBHs.

In Sec.~\ref{sec:abund} we outline the resultant constraints on the
mass fraction of the universe in PBHs formed from the collapse of
inflationary density perturbations, taking into account the formation,
due to critical collapse~\cite{nj}, of PBHs with a range of masses at
fixed horizon mass. Since we are assuming that supersymmetry is the
correct model of particle physics above $\sim 100$GeV we must apply
these constraints in the context of supersymmetric inflation
models. In Sec.~\ref{sec:susyinf} we discuss the constraints on
successful supersymmetric inflation models and calculate and review
PBH abundance constraints for two classes of inflation model; firstly those
where the reheat temperature after inflation is low, and then those
with a high reheat temperature and a subsequent period of thermal
inflation.

\section{Abundance of LSPs emitted by PBHs}
\label{sec:cond}
A PBH will emit LSPs if $T_{{\rm BH}}$ is greater than $m_{{\chi}}$
so that, using Eq.~(\ref{tbh}), if $M_{{\rm
BH}}<M_{{\rm 1}}$ where
\begin{equation}
M_{{\rm 1}} = \frac{1.06 \times 10^{13}{\rm GeV}}{m_{{\chi}}} 
          \,,
\end{equation}
then the PBH will emit LSPs throughout its evaporation. A PBH with
mass greater than $M_{{\rm 1}}$ will emit LSPs during the later stages
of its evaporation once its mass has fallen below $M_{{\rm 1}}$. The
the fraction of the PBH energy density eventually in LSPs will be
reduced by a factor of $M_{{\rm 1}}/ M_{{\rm BH}}$ relative to that
for PBHs with $M_{{\rm BH}} < M_{{\rm 1}}$.  In this paper we will
focus on the case of PBHs formed from the collapse of large
inflationary density perturbations. If the horizon mass at the time
the PBHs form is $M_{{\rm H}}$ then if $M_{{\rm H}}< M_{{\rm 1}}$ the
LSP emission will be dominated by PBHs with $M_{{\rm BH}} < M_{{\rm
1}}$. On extrapolating the constraint on the initial mass fraction in
PBHs which is found in Sec.~\ref{ti} (and incorporating the additional
weakening factor of $M_{{\rm BH}}/ M_{{\rm 1}}$) we can see that for
$M_{{\rm H}} > M_{{\rm 1}}$ the constraint from LSP emission will be
weaker than those from the effects of PBH evaporation on
nucleosynthesis. Therefore we will only calculate the constraints from
LSP emission by PBHs with $M_{{\rm BH}}<M_{{\rm 1}}$.

The PBH lifetime, in grams per second, is~\cite{MacG}
\begin{equation}
\tau(M_{{\rm BH}})= \frac{6.24 \times 10^{-27} M_{{\rm BH}}^3}
            {f(M_{{\rm BH}})} \,,
\end{equation}
so that the temperature at evaporation, $T_{{\rm evap}}=T_{{\rm Pl}} 
(t_{{\rm Pl}}/\tau(M_{{\rm BH}}))^{1/2}$, is given by
\begin{equation}
\label{Tevap}
T_{{\rm evap}} = \left( \frac{1.24 \times 10^{21}f(M_{{\rm 1}})}
               {M_{{\rm BH}}^{3}} \right)^{1/2} {\rm GeV} \,.
\end{equation}
For evaporation to occur after LSP freeze out ($T_{{\rm evap}}< T_{{\rm
f}}$), $M_{{\rm BH}}$ must be greater than $M_{{\rm 2}}$ where, using the
fact that $T_{{\rm f}} \sim m_{\chi}/25$,
\begin{equation}
M_{{\rm 2}}=\left(\frac{7.75 \times 10^{23} f(M_{{\rm 2}})}
               {m_{\chi}^{2}} \right)^{1/3} \,.
\end{equation}
A PBH will therefore evaporate after LSP freeze out and also emit LSPs
throughout its evaporation if its mass is in the range $M_{{\rm 2}}<
M_{{\rm BH}}< M_{{\rm 1}}$.

If at some initial time $t_{{\rm i}}$ the fraction of the total energy
density of the universe in PBHs which evaporate after freeze out
producing LSPs is $\beta_{\chi,{\rm i}}= \rho_{{\rm
pbh,i}}^{\chi}/\rho_{{\rm tot,i}}$, then immediately before
the PBHs evaporate
\begin{equation}
\label{ten}
\left( \frac{\rho_{{\rm pbh}}}{\rho_{{\rm rad}}}\right)_{{\rm evap}} =
      \frac{ \beta_{\chi,{\rm i}}}{1-\beta_{\chi,{\rm i}}} \frac{T_{{\rm i}}}
         {T_{{\rm evap}}} \,.
\end{equation}

 If the PBHs dominate the energy density of the
universe before they evaporate then, since the radiation emitted by
the PBH dominates the background radiation energy density, soon
after evaporation
the LSPs will come to dominate the energy density of the universe
soon afterwards. To avoid this we require
\begin{equation}
\label{evap1}
\frac{\beta_{{\rm i}}}{ 1- \beta_{{\rm i}}} < \frac{T_{{\rm evap}}}
            {T_{{\rm i}}} \,.
\end{equation}

The LSPs emitted will initially be relativistic with mean energy per
particle $3 T_{{\rm BH}}$ The ratio of the energy density in
LSPs to that in radiation is therefore constant
\begin{equation}
\left( \frac{\rho_{{\chi}}}{\rho_{{\rm rad}}} \right) = \epsilon_{\chi} \,,
\end{equation}
where $\epsilon_{\chi}$ is the fraction of the PBH mass energy
evaporated into LSPs, until the LSPs become non-relativistic at $T_{{\rm
nr}}$ where, since $\rho \propto T^4$ for relativistic fluids,
\begin{equation}
T_{{\rm nr}}=T_{{\rm evap}} \left( \frac{m_{\chi}}{3 T_{{\rm BH}}} 
         \right)^{1/4} \,.
\end{equation}
At any subsequent epoch, with temperature T, before 
matter-radiation equality 
\begin{eqnarray}
\left(\frac{\rho_{{\chi}}}{\rho_{{\rm rad}}} \right)& =&
     \left(\frac{\rho_{{\chi}}}{\rho_{{\rm rad}}} \right)_{{\rm nr}}
     \frac{T_{{\rm nr}}}{T} \nonumber \\
        & =& \left( \frac{\rho_{{\chi}}}{\rho_{{\rm rad}}} 
      \right)_{{\rm evap}} \frac{T_{{\rm evap}}}{T} \left( \frac{m_{\chi}}
      { 3 T_{{\rm BH}}} \right)^{1/4} \,.
\end{eqnarray}  
The fraction of the energy density of the universe in LSPs at
the present day is therefore given by 
\begin{eqnarray}
\label{ochi}
\Omega_{{\chi, 0}}&=&\Omega_{\chi, {\rm eq}} =2  \frac{ 
            T_{{\rm evap}}}{T_{{\rm eq}}} \left(\frac{m_{\chi}}
         {3T_{{\rm BH}}}\right)^{1/4}
           \left(\frac{\rho_{{\chi}}}  
            {\rho_{{\rm rad}} } \right)_{{\rm evap}} \nonumber \\
         &=& 2 \epsilon_{\chi} \frac{\beta_{\chi,{\rm i}}}
          {1- \beta_{\chi,{\rm i}}} 
        \frac{T_{{\rm i}}}{T_{{\rm eq}}} \left(\frac{m_{\chi}}
         {3T_{{\rm BH}}}\right)^{1/4} \,,
\end{eqnarray} 
where `eq' denotes the epoch of matter-radiation equality. This
relation can be inverted simply to obtain the constraint on
$\beta_{\chi,{\rm i}}$ from the requirement $\Omega_{\chi,0} < 1$, as
a function of $T_{{\rm i}}$. This constraint is independent of the
mechanism of PBH formation and whilst we have assumed that all PBHs
form at the same time $t_{{\rm i}}$ it would be simple to recalculate
the constraint allowing for PBH formation at a range of times.

The relevant mass range and the fraction of the mass of the PBH which
is evaporated into LSPs both depend on $f(M_{{\rm BH}})$. To fully
calculate $f(M_{{\rm BH}})$ as a function of $M_{{\rm BH}}$ we would
need to know the full mass spectrum of the sparticles. We can however
calculate $f(M_{{\rm BH}})$ in two limiting cases. If $T_{{\rm BH}}
\sim m_{{\chi}}$ then the LSP will be the only sparticle emitted
(along with photons, gravitons, gluons, the three lepton families and
six quark flavours) whilst if $T_{{\rm BH}} \gg m_{{\chi}}$ then all
the standard model particles (including the Higgs, Z and W bosons) and
their supersymmetric partners will be emitted.  The relativistic
contributions to $f(M_{{\rm BH}})$ per particle degree of freedom are
\cite{MacG}:
\begin{eqnarray}
f_{s=0}&=&0.267 ,\,\, f_{s=1}=0.060 \,, \nonumber \\
          f_{s=3/2}&=&0.020 ,\,\, f_{s=2}=0.007 \,, \nonumber \\ 
          f_{s=1/2}&=& \ 0.147 \,\, {\rm uncharged} \,, \nonumber \\ 
          f_{s=1/2}&=&  0.142 \,\, {\rm electrically \,\,charged}  \,.
\end{eqnarray}

\subsection{$T_{{\rm BH}} \sim m_{\chi}$}
\label{Tsim}
In this regime a PBH can emit the 3 lepton families, 6 quark
flavours\footnote{The LSP mass may be smaller than that of the top
quark in which case there may be a narrow range of PBH masses where
the LSP is emitted but the top quark is not, which would decrease the
value of $f(M_{{\rm BH}})$ calculated here by about $1.704$.}, the
graviton, the gluon and the LSP. The photon has 2 possible
polarisation states and the gluon has 48 degrees of freedom giving a
total of 50 $s=1$ degrees of freedom. Each type of neutrino has 2
degrees of freedom resulting in 6 neutral $s=1/2$ degrees of
freedom. The electron, muon and tau leptons each have 4 degrees of
freedom, whilst each quark has 12 degrees of freedom resulting in a
total of 84 charged $s=1/2$ states. Finally the graviton has $s=2$ and
4 degrees of freedom. The total contribution of standard model states
to $f(M_{{\rm BH}})$ is therefore $(50 \times 0.06)+ (6 \times 0.147)
+ (84 \times 0.142) + (4\times 0.007)=15.84 $. Finally, as outlined in
the introduction, the LSP is most likely to be the lightest neutralino
which is uncharged and has s=1/2, giving a final value of $f(M_{{\rm
BH}})=15.84 + (2 \times 0.147)=16.13$. The fraction of the total mass
which is evaporated into LSP is $\epsilon_{\chi}=0.294/16.13=0.018$.

\subsection{$T_{{\rm BH}} \gg m_{\chi}$}
If $T_{{\rm BH}}$ is much larger than $m_{\chi}$ all the standard
model particles and their supersymmetric partners will be emitted.  In
the Minimal Supersymmetric Standard Model there are 2 Higgs doublet
fields which give rise to 5 physical states: the charged Higgs, the
light scalar Higgs, the heavy scalar Higgs and the pseudoscalar Higgs
(see e.g.~Ref~\cite{susydm} for details). Along with the W and Z bosons
these states lead to 9 $s=1$ and 5 $s=0$ degrees of freedom in
addition to the standard model degrees of freedom considered in
Sec.~\ref{Tsim}, so that the total contribution of standard model
states to $f(M_{{\rm BH}})$ is 17.10

Each standard model (SM) degree of freedom had a supersymmetric (SUSY)
degree of freedom with $s=1/2,0,1/2$ and $3/2$ for standard model
states with $s=0,1/2,1$ and $2$ respectively. This leads to a grand
total of 95 $s=0$ (5 SM, 90 SUSY), 92 charged $s=1/2$ (84 SM, 8 SUSY),
30 uncharged $s=1/2$ (6 SM, 56 SUSY), 59 $s=1$ (all SM), 4 $s=3/2$
(all SUSY) and 4 $s=2$ (all SM) degrees of freedom, giving $f(M_{{\rm
BH}})=46.79$. The fraction of the total mass evaporated into
supersymmetric particles is $\epsilon_{{\rm SUSY}}=29.69/46.79=0.63$,
with $\epsilon_{{\chi}}=0.294/46.79=0.006$ of the total mass
evaporated directly into LSPs. Whilst the fraction of the PBH mass
evaporated directly into LSPs is small all the other supersymmetric
particles emitted will decay rapidly producing at least one additional
LSPs each so that we can set $\epsilon_{\chi} =\epsilon_{{\rm SUSY}}$
when estimate the fraction of the PBH mass ending up in the form of
LSPs.

\section{PBH formation from density perturbations}
\label{sec:abund}
In order for a PBH to be formed, a collapsing region must be large
enough to overcome the pressure force resisting its collapse as it
falls within its Schwarzschild radius. This occurs if the perturbation
is bigger than a critical size $\delta_{{\rm c}}$ at horizon
crossing. There also is an upper limit of $\delta < 1$ since a
perturbation which exceeded this value would correspond initially to a
separate closed universe~\cite{harr}. Analytic
calculations~\cite{dpcoll} find $\delta_{{\rm c}} \sim 1/3$ and
assume that all PBHs have mass roughly equal to the horizon mass at
the time they form, independent of the size of the perturbation.
Recent studies\cite{nj} of the evolution of density perturbations
have found that the mass of the PBH formed in fact depends on the size
of the perturbation:
\begin{equation}
\label{mbh}
M_{{\rm BH}}= k M_{{\rm H}} (\delta-\delta_{{\rm c}})^{\gamma} \,,
\end{equation}
where $\gamma \approx 0.37$ and $k$ and $\delta_{{\rm c}}$ are
constant for a given perturbation shape (for Mexican Hat shaped
fluctuations $k=2.85$ and $\delta_{{\rm c}}=0.67$).

In order to determine the number of PBHs formed on a given scale we must
smooth the density distribution using a window function, $W(kR)$. For
Gaussian distributed fluctuations the probability distribution of the
smoothed density field $p(\delta(M_{{\rm H}})$ is given by
\begin{eqnarray}
p(\delta(M_{{\rm H}})) \, {\rm d} \delta(M_{{\rm H}})& =& \frac{1}{ 
         \sqrt{2 \pi} \sigma(M_{{\rm H}})}  \nonumber \\
     &&	\times \exp{\left( - \frac{\delta^2(M_{{\rm H}})}
        {2 \sigma^2(M_{{\rm H}})}\right)} \, 
	{\rm d} \delta(M_{{\rm H}}) \,,
\end{eqnarray}
where $\sigma(M_{{\rm H}})$ is the mass variance evaluated at horizon
crossing. For power law spectra, $P(k) \propto k^{n}$, where $P(k) =
\langle |\delta_{{\bf k}} |^2 \rangle$ and $n$ is the spectral index,
$\sigma^2(M_{{\rm H}}) \propto M_{{\rm H}}^{(1-n)/4}$ during radiation
domination~\cite{n2}.  The formation of PBHs on a range of scales has
recently been studied \cite{bhscale} for power law power spectra and
for flat spectra with a spike on a given scale. In both cases it was
found that, in the limit where the number of PBHs formed is small
enough to satisfy the observational constraints on their abundance at
evaporation and at the present day, it can be assumed that all the
PBHs form at a single horizon mass. In particular if the power
spectrum is a power law with $n>1$, as is the case in tree-level
hybrid inflation models, all PBHs form at the
smallest horizon scale immediately after re-heating.

The initial mass fraction of the universe in PBHs with masses in the
range $M_2 < M_{{\rm BH}} < M_1$, which evaporate after LSP freeze out and
produce LSPs throughout their evaporation, is given by
\begin{eqnarray}
\label{bchi}
\beta_{\chi{\rm i}} & = & 
            \int_{\delta_2}^{\delta_1} \frac{M_{{\rm BH}}}{M_{{\rm H}}} 
            p(\delta(M_{{\rm H}})) {\rm d} \delta(M_{{\rm H}}) \nonumber \\
            & = & \int_{\delta_2}^{\delta_1} \frac{ k (\delta-
            \delta_{{\rm c}})^{\gamma}}{ \sqrt{2 \pi} \sigma(M_{{\rm H}})}
           \exp{\left(- \frac{\delta^2}{2 \sigma^2(M_{{\rm H}})} \right)}
           {\rm d}\delta(M_{{\rm H}}) \,,
\end{eqnarray}
where, using eq.~(\ref{mbh}),
\begin{equation}
\delta_{{\rm j}}= \delta_{{\rm c}} + \left( \frac{M_{j}}{k M_{{\rm H}}}
            \right)^{\gamma} :\,\, j=1,2 \,.
\end{equation}

Eq.~(\ref{bchi}) can be used to translate the constraints on
$\beta_{\chi, {\rm i}}$ into constraints on $\sigma(M_{{\rm H}})$, which in
turn can be used to constrain $n$~\cite{n1,KL,n2}. We can also find the
maximum fraction of the universe in PBHs of all mass, $\beta_{{\rm
i}}$:
\begin{equation}
\label{betatot}
\beta_{{\rm i}} =\int_{\delta_{{\rm c}}}^{1} \frac{ k (\delta-
            \delta_{{\rm c}})^{\gamma}}{ \sqrt{2 \pi} \sigma(M_{{\rm H}})}
           \exp{\left(- \frac{\delta^2}{2 \sigma^2(M_{{\rm H}})} \right)}
           {\rm d}\delta(M_{{\rm H}}) \,,
\end{equation}

\section{Constraints on supersymmetric inflation models}
\label{sec:susyinf}
Since we are assuming that supersymmetry is the correct model of
particle physics above $\sim 100$GeV then we must apply the
constraints on the abundance of PBHs formed from the collapse of large
inflationary density perturbation to models of inflation constructed
in the context of supersymmetry. Avoiding the overproduction of
various relic particles, which would alter the subsequent evolution of
the universe and wreck the successful predictions of the standard hot
big bang model, leads to constraints on supersymmetric inflation
models.

The gravitino (the supersymmetric partner of the graviton) has only
gravitational interactions and mass of order $100$ GeV, and will decay
after nucleosynthesis. The requirement that its decay products do not
destroy the successful predictions of big-bang nucleosynthesis places
limits on its abundance. Since the number density of gravitinos is
proportional to the reheat temperature after inflation $T_{{\rm RH}}$,
this leads to an upper limit on $T_{{\rm RH}}$~\cite{grav1}. The exact
limit depends on the gravitino mass, the available decay channels and
the baryon-to-photon ratio before the gravitino decays\cite{grav2}
but, very conservatively, $T_{{\rm RH}}$ must be less than $10^{9}$
GeV. Similarly in almost all theories in which supersymmetry is broken
at an intermediate scale there are scalar fields, known as
moduli~\footnote{Specific examples are the dilaton of string theory
and the massless gauge singlets of string compactifications or, in
general, any gauge singlet field responsible for SUSY breaking}, which
typically have the same mass and lifetime as the gravitino~\cite{mod}.
Avoiding their production after inflation requires $T_{{\rm RH}}$ to
be less than $10^{12}$ GeV.

There are several ways to avoid the gravitino and moduli
problems. Firstly inflation models can be constructed where inflation
occurs at a low energy scale~\cite{rsg} so that the reheat temperature
is automatically low enough to avoid these problems. However these
models in general require fine-tuning~\cite{randall}. Secondly the
reheat temperature can be sufficiently low if the inflaton is long
lived. For instance a model has been constructed~\cite{sarross}, using
a singlet field in a hidden sector, where inflation occurs at the
scale of the spontaneous breaking of the gauge symmetry which is of
order $10^{14}$GeV. In this case, since the inflaton has only
gravitational strength couplings, the reheat temperature is only of
order $10^{5}$ GeV.

There is however a generic solution to these problems in high energy
scale inflation models which relies on the properties of flaton fields
which also arise naturally in supersymmetric
theories~\cite{ti}. Flaton fields, have vacuum expectation values
$M \gg 10^3$~GeV, even though their mass $m$ is only of order
the supersymmetry scale, so that their potential is almost flat. In
the early universe these fields are held at zero by finite temperature
effects, with false vacuum energy density $V_{0} \sim m^2 M^2$. Once
the temperature falls below $V_{0}^{1/4}$, the false vacuum energy
density dominates the thermal energy density of the universe and
begins to drive a period of inflation known as thermal inflation. This
inflation continues until the temperature drops to $T \sim m$, at
which point thermal effects are no longer strong enough to anchor the
flaton in the false vacuum.  Taking $M\sim 10^{12}$~GeV gives
$V_{0}^{1/4} \sim 10^{7}$~GeV so that around $\ln(10^7/10^3) \sim 10$
$e$-foldings of thermal inflation occur, sufficient to dilute the
moduli and gravitinos existing before thermal inflation but small
enough to not affect the density perturbations generated during the
first period of inflation.

We examine two classes of inflation model; those with a low reheat
temperature and those with a high reheat temperature and a subsequent
period of thermal inflation.

\subsection{Low-reheat temperature}
\begin{figure}[t]
\centering
\leavevmode\epsfysize=6.3cm \epsfbox{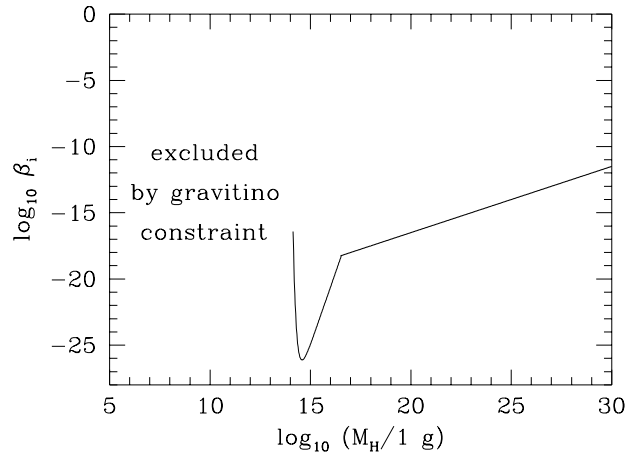}\\
\caption[blowt]{\label{blowt} The constraints on the initial mass
fraction of PBHs, $\beta_{{\rm i}}$, in inflation models with $T_{{\rm
RH}}< 10^{9}$ GeV .}
\end{figure}

If $T_{{\rm RH}} = 10^{9}$ GeV, so as to just satisfy the gravitino
constraint, then, using the relationship between horizon mass and
temperature in a radiation dominated universe:
\begin{equation}
M_{{\rm H}} \simeq 10^{18} \, {\rm g} \, \left( \frac{10^7 \, {\rm
	GeV}}{T} \right)^2 \,,
\end{equation}
the maximum horizon mass is $1 \times 10^{14}$g. Although when
critical collapse is taken into account PBHs with a range of masses
are formed at fixed horizon mass, the vast majority of PBHs have mass
within an order of magnitude of $M_{{\rm H}}$. From eq.~(\ref{ochi})
$\beta_{\chi,{\rm i}}$ must be less than $5 \times 10^{-17}$. However,
for $m_{\chi}=30$ GeV, only PBHs with $M_{{\rm BH}}< M_1 =3.5 \times
10^{11}{\rm g} \sim 0.0035 M_{{\rm H}}$ are hot enough to emit
LSPs. The fraction of PBHs which have mass this much smaller than the
horizon mass is negligible so that there is no resultant constraint on
$\beta_{{\rm i}}$. Similarly the constraints from the effect of the
products of PBH evaporation on nucleosynthesis only hold for $M_{{\rm
BH}} < 10^{13}$g. Therefore in a low reheat temperature inflation
model only the constraint from the present day density of PBHs and, if
$T_{{\rm RH }} > 5\times 10^{8}$ GeV, that on the abundance of PBHs
evaporating today hold. The constraints on $\beta_{{\rm i}}$ are shown
in fig.\ref{blowt}. The `U'-shaped dip arises from the limit on the
abundance of PBHs with mass $M_{{\rm BH}} \sim 5 \times 10^{14}$g
which are evaporating at the present day (as calculated by
Yokoyama~\cite{yok}) and the straight line from the limit on the
present day density of PBHs which have not evaporated.

The constraints on $\beta_{{\rm i}}$ can be translated into limits on
the spectral index of the density perturbations~\cite{n1,KL,n2} using
Eq.~(\ref{betatot}) and the scale dependence of $\sigma(M_{{\rm
H}})$:
\begin{equation}
\sigma(M_{{\rm H}})= \sigma(M_0)\left( \frac{M_{{\rm H}}}{M_{{\rm eq}}}
                \right)^{(1-n)/4} \left( \frac{M_{{\rm eq}}}{M_{{\rm 0}}}
                \right)^{(1-n)/6} \,,
\end{equation}
where $M_{0}$ and $M_{{\rm eq}}$ are the horizon masses at the present
day and at matter radiation equality and, using the COBE
normalisation, $\sigma(M_{0}=10^{56}{\rm g})=9.5 \times 10^{-5}$. In
models with $T_{{\rm RH}}< 10^{9}$ GeV, the tightest limit on $n$
($\sim 1.28$) arises from the constraint on PBHs which are evaporating
today. This is weaker than the limits found in Ref.~\cite{n2} firstly
because the tightest limits arise from the constraints on PBHs of mass
$M_{{\rm BH}} \sim 10^{10-11}$g which are not formed in low reheat
temperature inflation models. Also the analytically derived value of
$\delta_{{\rm c}}$ ($\sim 1/3$) used in those calculations is roughly
half the new value $\delta_{{\rm c}}=0.67$, determined from numerical
simulations~\cite{nj}, which is used in this paper.

\subsection{High-reheat temperature with thermal inflation occurring}
\label{ti}
The effects of a period of thermal inflation on the constraints on PBH
abundance were studied in Ref.~\cite{n2}. It is assumed that thermal
inflation commences at $T=10^{7}$ GeV and continues until $T=10^{3}$
GeV, when the flaton field rolls to its true vacuum state. Assuming
for simplicity that reheating is efficient then the universe is
reheated to $T=10^{7}$ GeV with the subsequent evolution of the
universe having its standard form. The duration of thermal inflation
is negligible compared to the PBH lifetime so that its main effect on
the PBHs is to dilute their density by a factor $\rho_{{\rm
i}}/\rho_{{\rm f}}= \left( a_{{\rm f}}/a_{{\rm i}} \right)^3 \sim
(10^4)^3 $ so that the constraints on $\beta_{{\rm i}}$ are weakened
by a factor of $\sim 10^{12}$.

\begin{figure}[t]
\centering
\leavevmode\epsfysize=6.3cm \epsfbox{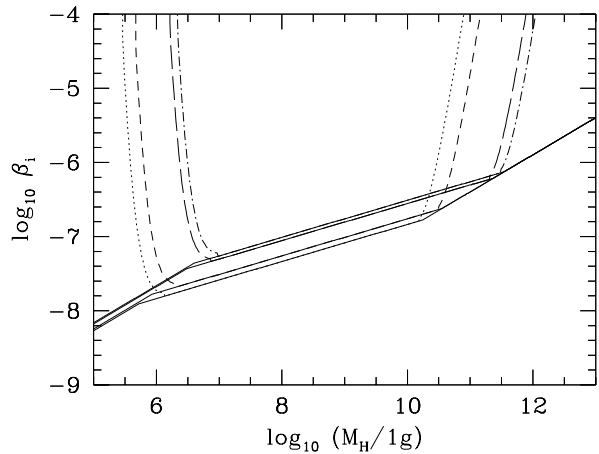}\\
\caption[traj1]{\label{betatotm} The constraints on the initial mass
fraction of PBHs from the present day density of LSPs, in
supersymmetric inflation models with a high reheat temperature and a
subsequent period of thermal inflation for $\epsilon_{\chi}=0.018$.
The solid lines shows the constraints on $\beta_{\chi, {\rm i}}$, the
fraction of the universe in PBHs which evaporate after freeze-out and
produce LSPs throughout their evaporation, for (from bottom to top)
$m_{\chi}= 600, 300, 45$ and $30$ GeV. The dotted, short dashed, long
dashed and dot-dashed lines show the maximum allowed fraction of the
universe in PBHs of all masses, $\beta_{{\rm i}}$, for $m_{\chi}= 600,
300, 45$ and $30$ GeV respectively.}
\end{figure}

\begin{figure}[t]
\centering \leavevmode\epsfysize=6.3cm \epsfbox{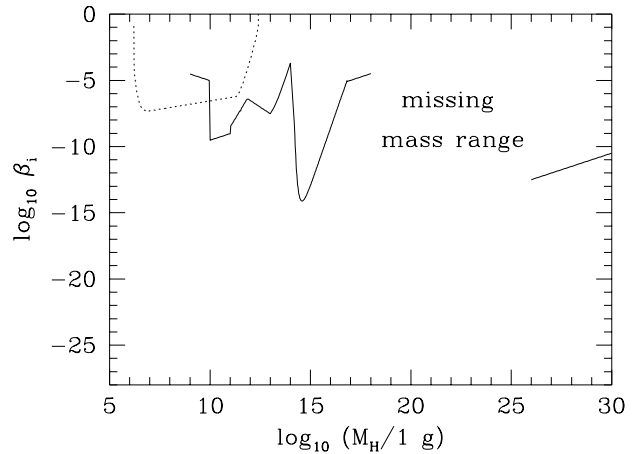}\\
\caption[traj1]{\label{fig3} A compilation of the constraints
on the initial mass fraction of PBHs in supersymmetric inflation
models with a high reheat temperature and a period of thermal
inflation. The dotted show the constraints from the present day
density of LSPs. From left to right the solid lines show the
constraints from: the effects of PBH evaporation after
nucleosynthesis, the abundance of PBHs evaporating at present, the
present day density of PBHs formed before and (on the far right) after
thermal inflation.}
\end{figure}

If thermal inflation occurs then the right hand sides of
eqs.(\ref{ten}--\ref{ochi}) are each multiplied
by a factor of $10^{-12}$ due to the dilution of the PBHs during
thermal inflation. Using the relation between temperature and horizon
mass $ T_{{\rm i}} = T_{{\rm Pl}} (m_{{\rm Pl}}/M_{{\rm H}})^{1/2}$ the
requirement that the PBHs do not dominate the universe at evaporation
becomes
\begin{equation}
\frac{\beta_{\chi,{\rm i}}}{1-\beta_{\chi,{\rm i}}}< 6.7 \times 10^{6} 
\left(\frac{M_{{\rm H}}}{M_{{\rm BH}}^3} \right)^{1/2} \,,
\end{equation}
and the limit from the present day density of LSPs becomes
\begin{equation}
\frac{\beta_{\chi,{\rm i}}}{1-\beta_{\chi,{\rm i}}}< \frac{5 \times 
          10^{11}}
        {\epsilon_{\chi}}\frac{T_{{\rm eq}}}{T_{{\rm Pl}}}
        \left(\frac{M_{{\rm H}}}{m_{{\rm Pl}}}\right)^{1/2}
        \left( \frac{3 \times 10^{13}}{m_{\chi} M_{{\rm BH}}}
        \right)^{1/4} \,.
\end{equation} 
To evaluate these constraints on $\beta_{\chi,{\rm i}}$ we neglect the
spread in PBHs masses, since the vast majority of PBHs will have the
same mass to within a factor of a few. If $M_{{\rm H}}$ is between
$M_{{\rm 2}}$ and $M_{{\rm 1}}$ then most of the PBHs will have
$M_{{\rm BH}} \sim M_{{\rm H}}$, whilst if $M_{{\rm H}}<M_{2}$ the
vast majority of the PBHs which evaporate after freeze out will have
$M_{{\rm BH}} \sim M_{2}$ and similarly if $M_{{\rm H}} > M_{1}$ most
of the LSP producing PBHs will have $M_{{\rm BH}} \sim M_{1}$. The
tightest constraint arises from the present day abundance of LSPs for
$M_{{\rm H}}< M_{1}$ and also for $M_{{\rm H}} > M_{1}$ if $M_{1}< 2
\times 10^{12}$g (which is the case for $m_{\chi} < 30$GeV as found
experimentally). We calculated the resulting constraints on
$\sigma(M_{{\rm H}})$, and hence $\beta_{{\rm i}}$, for 4 sample
values of the LSP mass: $m_{\chi}= $ 30, 45, 300 and 600 GeV, using
$\epsilon_{\chi} \sim 0.018$. As the horizon mass decreases the
fraction of the total number of PBHs formed which are heavy enough to
evaporate after freeze-out decreases.  Similarly as the horizon masses
increases the fraction of PBHs which are light enough to emit LSPs
decreases. Emission of LSPs after freeze-out constrains $\beta_{{\rm
i}}$ for $10^{6}{\rm g} \leq M_{{\rm H}} \leq 10^{10}$g.  The
constraints on $\beta_{\chi,{\rm i}}$ and $\beta_{{\rm i}}$ are shown
in Fig.~\ref{betatotm}. In Fig.~\ref{fig3} the constraints on
$\beta_{{\rm i}}$ for $m_{\chi}=45$GeV are shown, along with those
from the effects of PBH evaporation on the products of
nucleosynthesis~\footnote{The sharply rising line from $M_{{\rm H}}
\sim 10^{13}{\rm g}$ arises due to the constraint from entropy
production after nucleosynthesis on PBHs with mass $M_{{\rm BH}}<
10^{13}$g.}, the abundance of PBHs evaporating at present and the
present day density of PBHs as calculated in refs.~\cite{n2}
and~\cite{yok}. The missing mass range corresponds to comoving scales
which enter the horizon before thermal inflation, and are then pulled
back outside again during thermal inflation. Any new density
perturbations are expected to be small, since the energy scale of
thermal inflation is much lower than the original inflationary period,
and hence unable to form black holes when they re-enter the horizon
again after thermal inflation.

If thermal inflation occurs then the temperature at which a given
comoving scale crosses the Hubble radius is changed, whilst the
relation between horizon mass and temperature remains the
same so that the scale dependence of $\sigma(M_{{\rm H}})$
becomes~\cite{n2}
\begin{equation}
\sigma(M_{{\rm H}})= \sigma(M_0)\left[ \left( \frac{10^{7}}
               {10^{3}} \right)^2 
               \frac{M_{{\rm H}}}{M_{{\rm eq}}}
                \right]^{(1-n)/4} \left( \frac{M_{{\rm eq}}}{M_{{\rm 0}}}
                \right)^{(1-n)/6} \,.
\end{equation}
The limits on $n$ from the constraints on $\beta_{{\rm i}}$ due to LSP
emission range from $n<1.36$ at $M_{{\rm H}} \sim 10^{11}$ g to
$n<1.33$ at $M_{{\rm H}} \sim 5 \times 10^{5}$g. These limits are
slightly tighter than those from the nucleosynthesis constraints ($n
\sim 1.34$ to $1.37$ if the accurate value of $\delta_{{\rm c}} =0.67$
is used).

\section{Conclusions}
We have examined the consequences of supersymmetry for PBH abundance
constraints. PBHs with mass in the range $10^{5}{\rm g} <M_{{\rm BH}}
< 10^{11}$g evaporate after LSP freeze out and produce LSPs (and other
supersymmetric particles) throughout their evaporation. In most models
of supersymmetry the LSP is stable and the requirement that the
present day density of LSPs does not exceed the critical density
places a limit on the initial abundance of PBHs in this mass range.

We have studied the constraints on PBH abundance for two classes of
supersymmetric inflation model; those with a low reheat temperature
and those with a high reheat temperature and a subsequent period of
thermal inflation. If the reheat temperature is low the lightest PBHs
which can be produced will be evaporating at the present day and the
constraints from the present day density of LSPs and the effects of
evaporation on the products of nucleosynthesis, which provide the
tightest limits on $n$, do not apply. For models with a high
reheat temperature and a subsequent period of thermal inflation the
constraints from the present day density of LSPs provides the tightest
limit on $n$.

\section*{Acknowledgements}

A.M.G.~was supported by PPARC and acknowledges use of the Starlink
computer system at QMW. A.M.G~thanks Lev Kofman, Andrew Liddle and
David Lyth for useful discussions.


\begin{references}
\bibitem{dpcoll} B J. Carr
           and S. W. Hawking, Mon. Not. R. Astron. Soc. {\bf 168}, 399 (1974).
\bibitem{form} S. W. Hawking, I. Moss and J. Stewart, Phys. Rev. D {\bf 26},
          2681 (1981); M. Crawford and
          D. N. Schramm, Nature {\bf 298}, 538 (1982); A. G. Polnarev and R. 
          Zembowicz, Astron, Zh. {\bf 58}, 706 (1988); D. La and P. J. 
          Steinhardt Phys. Lett. B {\bf 220}, 375 (1989); S. W. Hawking, Phys. 
          Lett. B {\bf 231}, 237 (1989); S. D. H. Hsu, Phys. 
          Lett. B {\bf 251}, 343 (1990).
\bibitem{MacG} J. H. MacGibbon and B. R. Weber, Phys. Rev. D {\bf 41},
          3052 (1990); J. H. MacGibbon, Phys. Rev. D {\bf 44}, 376 (1991).
\bibitem{bnuc}Ya. B. Zel'dovich, A. A. Starobinsky, M. Y. 
         Khlopov and V. M. Chechetkin, Pis'ma Astron. Zh. {\bf 3}
         , 308 (1977) [Sov Astron. Lett. {\bf 22}, 110 (1977);
        B. V. Vainer, D. V. Dryzhakova and P. D. Nasselskii,
         Pis'ma Astron. Zh. {\bf 4}, 344 (1978) [Sov. Astron. Lett
        {\bf 4},185 (1978);S. Mujana and K. Sato,
         Prog. Theor. Phys. {\bf 59},1012 (1978); B. V. Vainer and
         P. D. Nasselskii, Astron. Zh {\bf 55}, 231 (1978)
         [Sov. Astron. {\bf 22}, 138 (1978); D. Lindley, Mon. Not. R. 
         Astron. Soc. {\bf 193}, 593 (1980)
\bibitem{bgam}J. H. MacGibbon, Nature {\bf 320}, 308 (1987);
         J. H. MacGibbon and B. Carr, Astrophys. J. {\bf 371}, 447
         (1991), H. I. Kim, C. H. Lee and J. H. MacGibbon, Phys. Rev.
        {\bf D}, 063007 (1999).
\bibitem{susy} see e.g. J. Wess and J. Bagger, {\em Supersymmetry and 
          Supergravity} (Princeton University Press, Princeton, 1983). 
\bibitem{gh} L. Susskind, Phys. Rev. D {\bf 20}, 2619 (1979).
\bibitem{unif} U. Amaldi, W. de Boer and H. Furstenau, Phys. Lett. B 
            {\bf 260}, 447 (1991); C. Giunti, C. W. Kim and U. W. Lee,
            Mod. Phys. Lett. {\bf 16}, 1745 (1991); J. Ellis, S. Kelley and
            D. V. Nanopoulos, Phys. Lett. B {\bf 249}, 441 (1990); B 
           {\bf 260}, 131 (1991); P. Langacker and M. -X. Luc, Phys. Rev. 
           D {\bf 44}, 817 (1991).
\bibitem{wdm} G. Giuduce and R. Rattazzi, preprint hep-ph/9801271.
\bibitem{neutweak} J. Ellis, J. S. Hagelin, D. V. Nanopoulos, K. A. Olive,
         M. Srednicki, Nucl Phys. {\bf B} 238, 453 (1984). 
\bibitem{abund} J. Rich, M. Spiro and J. Lloyd-Owen, Phys. Rep. {\bf 151},
         239 (1987); P. F. Smith, Contemp. Phys. {\bf 29}, 159 (1998);
         T. K. Hemmick et. al. Phys. Rev. D {\bf 41}, 1940 (1998).
\bibitem{m1} LEP Experiments Committee meeting, Nov. 12th 1998, 
          http://www.cern.ch/Committees/LEPC/minutes \\
         /LEPC50.html
\bibitem{m2} J. Ellis, T. Falk, K. Olive and M. Schmitt, Phys. Lett. B
          {\bf 413}, 355 (1997).
\bibitem{susydm} G. Jungman, M. Kamionkowski and K. Griest,
          Phys. Rept. {\bf 267}, 195 (1996).
\bibitem{m3} K. A. Olive and M. Srednicki, Phys. Lett. B {\bf 230}, 78 (1989)
          and Nucl. Phys. B {\bf 355}, 208 (1991). K. Griest, M. Kamionkowski
          and M. S. Turner, Phys. Rev. D {\bf 41}, 3565 (1990). 
\bibitem{m4} J. Ellis, T. Falk and K. A. Olive, Phys. Lett. B {\bf 444},
            367 (1998).
\bibitem{nj} J. C. Niemeyer and K. Jedamzik, Phys. Rev. Lett. {\bf 80}, 
           5481 (1998); J. C. Niemeyer and K. Jedamzik, Phys. Rev. D 
           {\bf 59}, 124013 (1999).
\bibitem{harr} E. R. Harrison, Phys. Rev. D {\bf 1} 2726 (1970).
\bibitem{n2} A. M. Green and A. R. Liddle, Phys. Rev. D {\bf 54}, 6166 (1997).
\bibitem{bhscale} A. M. Green and A. R. Liddle, pre-print astro-ph/9901268.
\bibitem{n1} B. J. Carr, J. H. Gilbert and J. E. Lidsey, Phys. Rev. D 
            {\bf 50}, 4853 (1994).
\bibitem{KL} H. I. Kim and C. H. Lee, Phys. Rev. D {\bf 54}, 6001 (1996).
\bibitem{grav1} J. Ellis, D. Nanopoulos and M. Quiros, Phys. Lett. B
             {\bf 174}, 176 (1986).
\bibitem{grav2} T. Moroi, PhD Thesis Tohoku (1995), hep-ph/9503210.
\bibitem{mod} G. D. Coughlan, W. Fischler, E. W. Kolb, S. Raby and
	G. G. Ross, Phys. Lett. {\bf 131B}, 59 (1983);  J. Ellis,
	D.V. Nanopoulos and M. Quiros, Phys. Lett. {\bf 174B}, 176
	(1986);
        B. de Carlos, J. A. Casas, F. Quevedo and E. Roulet,
	Phys. Lett. B {\bf 318}, 447 (1993);
        T. Banks, D. B. Kaplan
	and A. E. Nelson, Phys.  Rev. D {\bf 49}, 779 (1994);
         L. Randall and S. Thomas, Nucl. Phys. B {\bf 449}, 229 (1995);
        T. Banks, M. Berkooz and P. J. Steinhardt, Phys. Rev. D {\bf
	52}, 705 (1995); M. Dine, L. Randall and S. Thomas,
	Phys. Rev. Lett. {\bf 75}, 398 (1995).
\bibitem{rsg} L. Randall, M. Solja\u{c}i\'{c} and A. Guth, 
         Nucl. Phys. B {\bf 472}, 377 (1996).
\bibitem{randall} L. Randall, in {\em Perspectives on Higgs Physics II,}
         ed. G. L. Kane, World Scientific, Singapore.
\bibitem{sarross} G. G. Ross and S. Sarkar, Nucl. Phys. B {\bf 461}, 597 
        (1996).
\bibitem{ti}D. H. Lyth and E. D. Stewart,
	Phys. Rev. Lett. {\bf75}, 201 (1995);
         D. H. Lyth and E. D. Stewart, Phys. Rev. D {\bf
	53}, 1784 (1996).
\bibitem{yok} J. Yokoyama, Phys. Rev. D. {\bf 58}, 107502 (1998).


\end{references}
\end{document}